\documentclass{PoS}
\usepackage[utf8]{inputenc}       
\usepackage[english]{babel} 
\usepackage{amssymb,amsmath,latexsym,enumerate}
\usepackage{cite}  
\usepackage{graphicx, fontenc, times, mathptmx }
\usepackage{indentfirst}      
\usepackage{array,longtable,lscape}

\title{3D numerical study of an anisotropic heat transfer in outer layers of magnetized neutron stars}

\ShortTitle{3D anisotropic heat transfer in neutron stars}

\author{\speaker{Ilya A. Kondratyev}$^{ab}$, Segrey G. Moiseenko$^a$, Gennady S. Bisnovatyi-Kogan$^{a,c}$, Maria V. Glushikhina$^a$ \\
        \llap{$^a$} Space Reseach Institute RAS\\
        Profsoyuznaya st. 84/32, Moscow, Russia, 117997\\
        \llap{$^b$}National Research University Higher School of Economics, Department of Physics \\
        Staraya Basmannaya st. 21/4 b.5, Moscow, Russia, 105066 \\
        \llap{$^c$}National Reseach Nuclear University MEPhI \\
         Kashirskoe sh., 31, Moscow, Russia, 115409\\
        E-mail: \email{mrkondratyev95@gmail.com}, \email{moiseenko@iki.rssi.ru}, \email{gkogan@iki.rssi.ru}, \email{m.glushikhina@iki.rssi.ru} }

\abstract{Periodic changes in a thermal soft X-ray flux of a rotating neutron star indicate a non-uniform distribution of the surface temperature. A possible cause of this phenomenon is a suppression of the heat flux across the magnetic field lines in a crust and an envelope of magnetized neutron stars. In this paper we study three-dimensional effects, associated with non-axisymmetric magnetic fields in neutron stars. We calculate the surface temperature distribution by solving numerically a three dimensional heat transfer equation in a magnetized neutron star crust. We adopt an anisotropic (tensorial) electron thermal conductivity coefficient, which is derived as an analytical solution of the Boltzmann equation with a Chapman-Enskog method. To calculate the surface temperature distribution, we construct a local one-dimensional plane-parallel model ("Ts-Tb"-relationship) of a magnetized neutron star envelope. We then use it as an outer boundary condition for the three-dimensional problem in the crust to find the self-consistent solution. To study possible observational manifestations from anisotropic temperature distributions we calculate light curves with a composite black-body model. Our calculations show, that a non-axisymmetric magnetic field distribution can lead to the irregular non-sinusoidal shape of a pulse profile as well as in some cases a significant amplification of pulsations of the thermal flux in comparison to the pure-dipolar magnetic field configurations.
}

\FullConference{High Energy Phenomena in Relativistic Outflows VII - HEPRO VII\\
		9-12 July 2019\\
		Facultat de Física, Universitat de Barcelona, Spain}

\begin{document}

\section{Introduction}
Neutron star (NS) surface magnetic fields can reach $\sim 10^{12-13}$G and even more on their surfaces. One of the possible ways to study surface magnetic fields is to observe thermal radiation in a soft X-ray band \cite{obs1, obs2}. Periodic changes of spectra from some X-ray dim isolated NSs (XDINSs) may indicate their non-uniform surface temperature distribution. Such heterogeneities are explained \cite{fi76} by an anisotropic thermal conductivity of degenerate matter in presence of a strong magnetic field, because a heat flux is suppressed across the magnetic field lines in the outer layers of such neutron star. Comparison results of simulations of the heat transfer processes to observational data can help to determine magnetic field structure on the NS surface and its interior.

Outer layers of the NS consist of a plasma of degenerate electrons and almost non-degenerate non-relativistic ions. We separate outer layers into the crust ($10^{10}<\rho< 2\cdot10^{14}$ g/cm$^3$), where matter can form a state of the Coulomb crystal or liquid, and an outer envelope ($\rho< 10^{10}$ g/cm$^3$). The thermal conductivity is mostly determined by electrons, and by radiation in a thin non-degenerate layer of the envelope near the surface. A degree of the heat flux suppression across the magnetic field is determined by a magnetization parameter $\omega\tau$, where $\tau$ is the average time between electron-nucleon collisions, and $\omega = eB/m_e^*c$ is the cyclotron plasma frequency; $m_e^* = m_e\sqrt{1 + {p_{fe}}^2 /m_ec^2}$ is the effective electron mass,  $p_{fe} = \hbar(3\pi^2n_e)^{1/3}$ is the electron Fermi momentum, $e$ is the electron charge, $c$ is the speed of light, $\hbar$ is the reduced Planck constant, $B$ is a local value of the magnetic induction.

In this work we look for a stationary solution of a 3D heat transfer equation with anisotropic thermal conductivity \cite{bisn1}. Because of a sufficient nonlinearity of a radiative boundary condition ($F\sim T^4$, where $F$ is a heat flux, $T$ is a temperature), physical parameters also drastically change through the thin crustal and envelope regions. In such case a full 3D approach can not be applied to the problem. Fortunately,  the heat flux in the envelope is mostly radial, and the temperature distribution in an envelope region can be calculated separately from the crust but self-consistently with it \cite{heat1}. We have built a local one-dimensional model of a thermal structure of the outer envelope. 

In 2D approach the problem of finding the stationary temperature distribution in the NS was solved in many works with different magnetic fields and microphysical input (e.g. \cite{heat1,heat2}, and \cite{pot2015} for a review).
\section{Physical model}

\subsection{Heat transfer in a magnetic field}

Temperature distribution is defined by the heat transfer equation:
\begin{equation}
C\frac{\partial T }{\partial t} = \nabla\cdot\kappa\cdot\nabla T + f,
\end{equation}
where $C$ is the heat capacity, $\kappa$ is the thermal conductivity tensor, $f$ is defined by heat sources and sinks (neutrino emission, Joule heating, etc.). We look for a stationary solution ($\frac{\partial T }{\partial t} = 0$) and also assume an absence of sources and sinks and consider $f=0$.

Thermal conductivity tensor $\kappa$ for strongly degenerate electrons in the magnetic field was obtained  in~\cite{bisn1, bisn2} from the solution of a Boltzmann equation with a Chapman-Enskog method. This tensor includes the effect heat fluxes along and across the magnetic field as well as the Hall heat flux. In Cartesian coordinates it is written as follows \cite{bisn1}
\begin{equation}
\begin{cases}
{\kappa}_{ij} = \frac{k_B^2Tn_e}{m_e^*}\tau\big(\kappa^{(1)}\delta_{ij} + \kappa^{(2)}\varepsilon_{ijk}\frac{B_k}{B} + \kappa^{(3)}\frac{B_iB_j}{B^2}\big) \\
\kappa^{(1)} = \frac{5\pi^2}{6}\big(\frac{1}{1 + (\omega\tau)^2} - \frac{6}{5}\frac{(\omega\tau)^2}{(1 + (\omega\tau)^2)^2} \big) \\
\kappa^{(2)} = -\frac{4\pi^2}{3}\omega\tau\big(\frac{1}{1 + (\omega\tau)^2} - \frac{3}{4}\frac{(\omega\tau)^2}{(1 + (\omega\tau)^2)^2} \big) \\
\kappa^{(3)} = \frac{5\pi^2}{6}(\omega\tau)^2\big(\frac{1}{1 + (\omega\tau)^2} + \frac{6}{5}\frac{1}{(1 + (\omega\tau)^2)^2} \big) 
\end{cases}
\label{heatcond}
\end{equation} 
where $n_e = \frac{\rho Z}{Am_u}$ is the electron number density, $\tau = \frac{3}{32\pi^2}\frac{h^3}{m_e^*Ze^4\Lambda}$ is the average time between electron-nucleon collisions, $k_B$ is the Boltzmann constant, $h$ is the Planck constant, $\Lambda$ is the Coulomb logarithm. Parameter $\omega\tau$ changes drastically in the crust and the envelope of the NS: at the density $\rho \sim 10^{10}$ g/cm$^3$, the value of the parameter $\omega\tau \sim 1$ when the magnitude of the magnetic field induction $B \sim 10^{13}$G. Approximately $\omega\tau \sim B/\rho^{2/3}$ in the crust for the degenerate electron gas. As it follows from \eqref{heatcond}, heat conductivity coefficients, aligned and transverse to the magnetic field, can be written in the following form:
\begin{equation}
\begin{cases}
\kappa_{e\parallel} = \frac{k_B^2Tn_e}{m_e^*}\tau\big(\kappa^{(1)} + \kappa^{(3)}\big),\\
\kappa_{e\perp} = \frac{k_B^2Tn_e}{m_e^*}\tau\kappa^{(1)}.
\end{cases}
\label{heatfl}
\end{equation}

Throughout this work the magnetic field is assumed to be a sum of dipolar and quadrupolar components
\begin{equation}
{\bf B} = {\bf B}_{\text{dip}} + {\bf B}_{\text{quad}},
\label{mfield}
\end{equation}
where the dipolar field strength is 
\begin{equation}
{\bf B}_{\text{dip}} = \frac{B_{pd}R^3_{NS}}{2}\frac{3({\bf d}\cdot {\bf r})\,{\bf r} - {\bf d}\,r^2}{r^5},
\label{dipole}
\end{equation}
and the quadrupolar field is represented by the following expression:
\begin{equation}
{\bf B}_{\text{quad}} = B_{pq}R^4_{NS}\bigg(\frac{r^2 - 5z^2}{2r^7}{\bf r} - \frac{{\bf e_z}z}{r^5}\bigg),
\label{quadrupole}
\end{equation}

where $R_{NS}$ is a NS radius, ${\bf d}$ is a unit vector aligned to magnetic dipole in the centre of a star, ${\bf e_z}$ is the unit vector aligned to the z-axis, and $B_{pd}$, $B_{pq}$ are polar magnetic field inductions of the the dipolar and quadrupolar fields correspondingly.

The degree of the heat flux suppression across the field ($\kappa^{(1)}$ in \eqref{heatcond} and \eqref{heatfl}) is stronger, than in previous works \cite{fi76, YakUrp}, where the relation between heat fluxes along and across the magnetic field is $\frac{F_\parallel}{F_\perp} = 1 + (\omega\tau)^2$.

\subsection{Equation of state}

Density explicitly appears in thermal conductivity \eqref{heatcond}. We further build a NS model solving Tolman-Oppenheimer-Volkoff equations for the hydrostatic equilibrium to obtain a density profile in the crust. For the NS interior we used a unified moderately stiff equation of state SLy4 \cite{EOS}, which is based on the microscopic calculations with an effective nuclear potential \cite{nucl1}. We have chosen central density $\rho_c= 10^{15}$ g/cm$^3$. NS mass is $M_{NS} = 1.42\,M_{\odot}$, where $M_{\odot}$ is a solar mass, and inner and outer radii of NS crust are $R_{in} = 10.594$ km at $\rho = \rho_{in} = 2\cdot10^{14}$ g/cm$^3$ and $R_{out} = R_{NS} = 11.618$ km at $\rho = 10^{10}$ g/cm$^3$ respectively.

For the outer envelope (see next subsection) we have assumed equation of state for non-quantizing ideal plasma with non-degenerate non-relativistic ions (nuclei) and degenerate relativistic electrons:
\begin{displaymath}
P = P_{id}^{(N)}+P_{id}^{(e)},
\end{displaymath}
where $P_{id}^{(N)} = n_N k_BT$ is the ion pressure, and the pressure of the electrons of an arbitrary degree of degeneracy can be written in terms of Fermi-Dirac integrals \cite{blin, PRE}:
\begin{equation}
P_{id}^{(e)} = \frac{(2m_e)^{3/2}}{3\pi^2\hbar^3\beta^{5/2}}\big(I_{3/2}(\chi,\tau) + \frac{\tau}{2}I_{5/2}(\chi,\tau)\big),
\label{Pe}
\end{equation}
where $\beta = (k_BT)^{-1}$, $\chi = \beta\mu_{id}^{(e)}$, and $\mu_{id}^{(e)}$ is the electron chemical potential, divided by $k_BT$, $\tau = (\beta m_ec^2)^{-1}$, and the Fermi-Dirac integral:
\begin{equation}
I_{\nu}(\chi,\tau) = \int_0^{\infty}\frac{u^\nu \sqrt{1 + \tau u/2}}{\exp(u-\chi) + 1}du,
\label{fermi}
\end{equation}
where $u = \beta m_ec^2(\sqrt{1+\frac{p^2c^2}{m_e^2c^4}}-1)$, and $p$ is the electron momentum. In most of the envelope and the crust the electron gas is strongly degenerate, and pressure can written analytically (e.g. \cite{LL1980}):
\begin{equation}
P_{sd}^{(e)} = \frac{m_e^4c^5}{32\pi^2\hbar^3}\bigg(\frac{1}{3}\sinh \xi -\frac{8}{3}\sinh\frac{\xi}{2} + \xi\bigg),
\label{Psd}
\end{equation}
where $\xi = 4\sinh^{-1}\frac{(3\pi^2n_e)^{1/3}\hbar}{m_e^2c^2}$. In non-degenerate layer near the surface in an ideal plasma approximation the electron pressure \eqref{Pe} approaches that of an ideal gas $P_{nd}^{(e)} =  n_e k_BT$.
We adopt analytical approximations for the Fermi-Dirac integrals from \cite{blin, PRE}.

\subsection{Thermal structure of an outer envelope}
Combining one-dimensional models for the NS outer envelope and 2D calculations of the heat transfer in the NS crust is any established way to investigate thermal properties of the NS (see \cite{pot2015} for a review). We followed the same line in our research.

The outer envelope of the NS is a thin layer ($\sim10-100$ metres) of plasma with degenerate electrons and non-degenerate ions. The temperature decreases by 2-3 orders of magnitude radially across this envelope, while surface temperature variations are within a factor of 10, so in the first approximation, the heat flux can be assumed to be purely radial throughout the envelope, and its local value is determined by a local surface temperature, so that $F_s = \sigma T_s^4$, $\sigma$ is the Stephan-Boltzmann constant. In such approach the thermal structure equation for the envelope reads \cite{G1983,PY01}:

\begin{equation}
\frac{dT}{dP} = \frac{3K}{16g_s}\frac{T_s^4}{T^3},
\label{TSE}
\end{equation}
where $T_s$ is the local surface temperature, $K = K(B, \theta_B, T,\rho)$ is the effective opacity, $\theta_B$ is the magnetic field inclination angle to a normal to the surface, $g_s = GM_{NS}/(R_{NS}^2\sqrt{1 - r_g/R_{NS}})$ is the surface gravity acceleration, where $G$ is the gravitational constant, and $r_g$ is the gravitational radius of the NS. For the effective opacity we assume the electron heat transfer \eqref{heatfl} and radiative one for free-free and bound-free transitions and electron Thomson scattering with taking into account the degeneracy of the electrons \cite{kondmoisBKGl}. Due to additivity of the thermal conductivity coefficient for electron and radiative processes the exact effective opacity reads as follows (e.g. \cite{bisn3}):
\begin{equation}
K^{-1} = K_e^{-1} + K_r^{-1},
\label{opacity}
\end{equation}
where $K_e = K_{ff} + K_{bf} + K_{Th}$ is the radiative opacity, and $K_e = 16\sigma T^3/3\kappa_e\rho$ is the electron opacity. A dependence of the thermal conductivity on the field inclination angle $\theta_B$ is given by the expression $\kappa = \kappa_{\parallel}\cos^2\theta_B + \kappa_{\perp}\sin^2\theta_B$ from \cite{GH1983}.

Equation \eqref{TSE} can be solved as a Cauchy problem for the given values of the surface temperature $T_s$ and surface pressure $P_s$, which is calculated from the Eddington approximation $P_s\approx \frac{2g_s}{3K(B, \theta_B, T_s,\rho_s)}$ \cite{bisn3}, using equation of state $P_s = P(\rho_s, T_s)$ to obtain $\rho_s$, where $\rho_s$ is the density at the NS surface. We solve this equation from the NS surface to the bottom of the envelope (crust-envelope boundary) at $\rho_b = 10^{10}$ g/cm$^3$ and obtain a temperature $T_b$ at this density. Solving equation \eqref{TSE} numerically for different magnetic fields, inclination angles and surface temperatures leads to a so-called $T_s-T_b$-relationship, tabulated or analytically approximated function, which relates the surface temperature $T_s$ with the temperature at the boundary between the crust and the envelope $T_b$. Results of several calculations are presented on Table 1, more results and a detailed discussion can be found in \cite{kondmoisBKGl}. We use the tabulated $T_s-T_b$-relationship as a boundary condition for the heat transfer equation in the crust (see next section).

\begin{table}[!htp]
	
	\begin{center}	
		\begin{tabular}{ | c | c | c | c | }	
			\hline
			$\lg B_{pd}$ & $11$ & $12$ & $13$ \\ \hline
			$T_s(\theta=0)/10^6K$ & 1.02 & 1.03 & 1.16 \\
			$T_s(\theta=\pi/2)/10^6K$ & 0.71 & 0.35 & 0.18 \\
			$T_{s\parallel}/T_{s\perp}$ & 1.43 & 2.94 & 6.44 \\
			\hline
		\end{tabular}
	\end{center}
	\caption{ Surface temperature on a NS magnetic pole $(\theta_B=0)$ and on an equator $(\theta_B=\pi/2)$ and their relations for $T_b=10^8K$ and the dipolar magnetic field \eqref{dipole} with induction $B_{pd}$ on the pole. }
\end{table}

\section{A boundary-value problem}
We assume, that the heat flux radiates from the surface of the NS, and the temperature on the inner radius of the crust $T_{core}$ is constant through the core because of the sufficiently large value of the heat capacity. This temperature decreases with time as the core cools very slowly. After fast neutrino cooling stage, thermal evolution of the NS can be considered as a sequence of cooling models with the static temperature distributions.

The value of the surface temperature $T_s$ is determined by the solution of a boundary-value problem for the stationary heat transfer equation in a spherical layer with a given temperature $T_{core}$ on the inner boundary, and the radiative blackbody boundary condition on the outer bound, which is described by the Stephan-Boltzmann law $F_s = \sigma T_s^4$. In the outer envelope, the radial heat flux $F_s$ is assumed to be constant. For this reason, this heat flux on the outer boundary of the crust (inner boundary of the envelope), is equal to the radial part of the heat flux calculated from the crust via the solution of the  heat transfer equation. Boundary conditions read as

\begin{equation}
T |_{in} = T_{core}, \quad \kappa ({\bf B}, \rho, T)\nabla_r{T} + F_s |_{out}= 0,
\label{bcore}
\end{equation}
where index $"in"$ corresponds to the value of the temperature of the inner boundary of the crust with $r = R_{in}$, and index $"out"$ corresponds to the outer one with $r = R_{out}$. Continuity of the temperature at the crust-envelope boundary is used with the $T_s-T_b$-relationship for different ($B,\, \theta_B$), also $T_b = T_{out}$. As a result, we obtain unambiguous dependence $T_s(T_{out},\, B,\, \theta_B)$, which determines the surface temperature distribution in the magnetized NS. In spherical layer $R_{in}\le r \le R_{out}$ we solve boundary-value problem for the heat transfer equation 
\begin{equation}
\nabla\cdot \kappa ({\bf B}, \rho, T)\cdot\nabla{T} = 0
\label{heattr}
\end{equation}
with the boundary conditions \eqref{bcore}. This problem is solved with our extension of the basic operators methods on an unstructured tetrahedral mesh \cite{kondmois,kondmoisBKGl}.

\section{Results}
\subsection{Temperature distributions}
We calculate self-consistently the temperature distributions inside the volume of the crust and on the surface of the NS, as well as thermal light curves for them. The latter are calculated adopting a composite blackbody model. More results and discussion can be found in our forthcoming paper \cite{3dheattr}. The inclusion of the quadrupolar field \eqref{quadrupole} adds two more physical parameters to the problem: the relation between the quadrupolar and dipolar field inductions $\beta = B_{pq}/B_{pd}$ and the angle between their magnetic axes $\Theta_b$.
\begin{figure}[!htp]
	\centering
	\includegraphics[width=7.5cm,height=7.0cm]{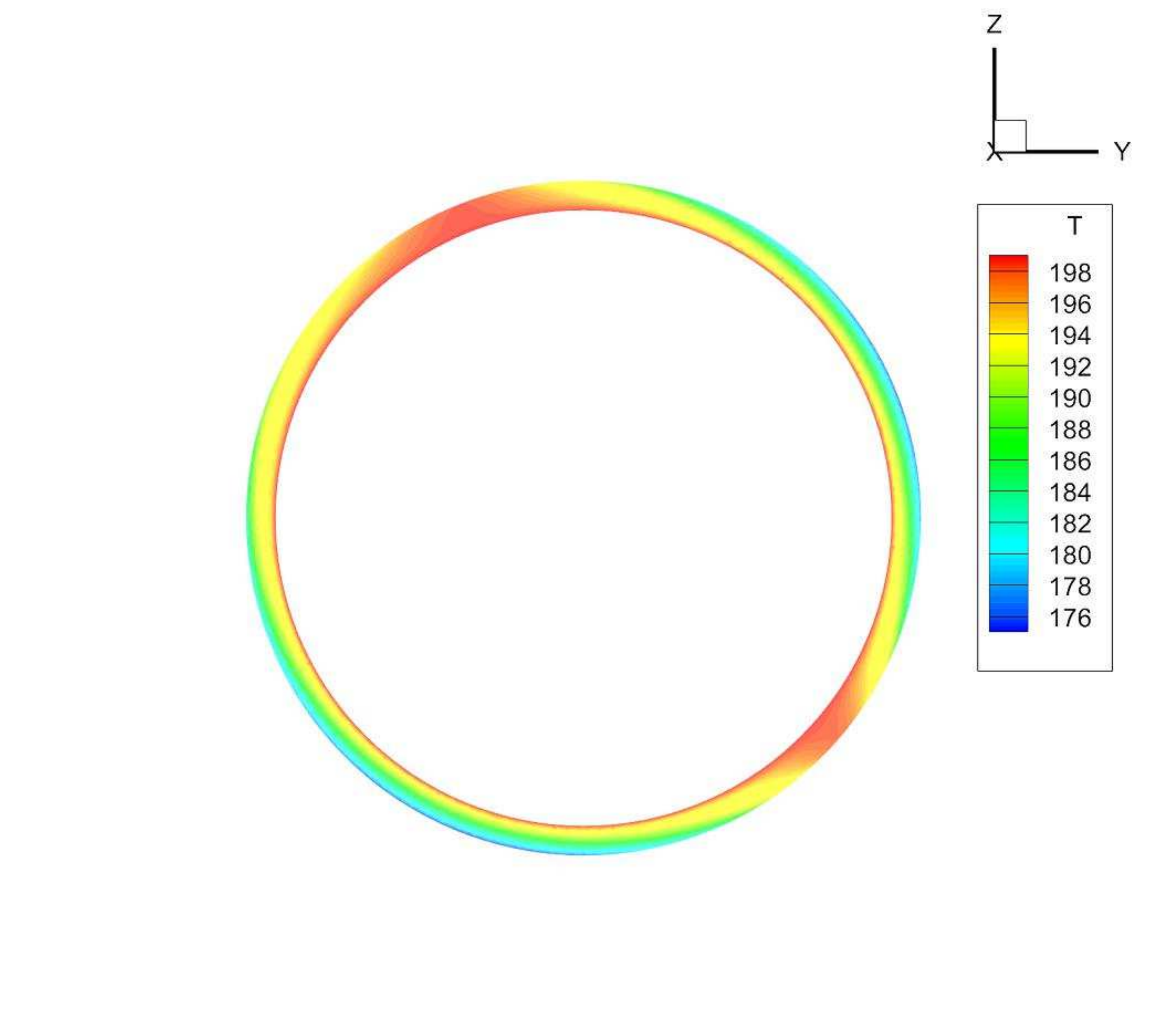}
	\includegraphics[width=7.5cm,height=7.0cm]{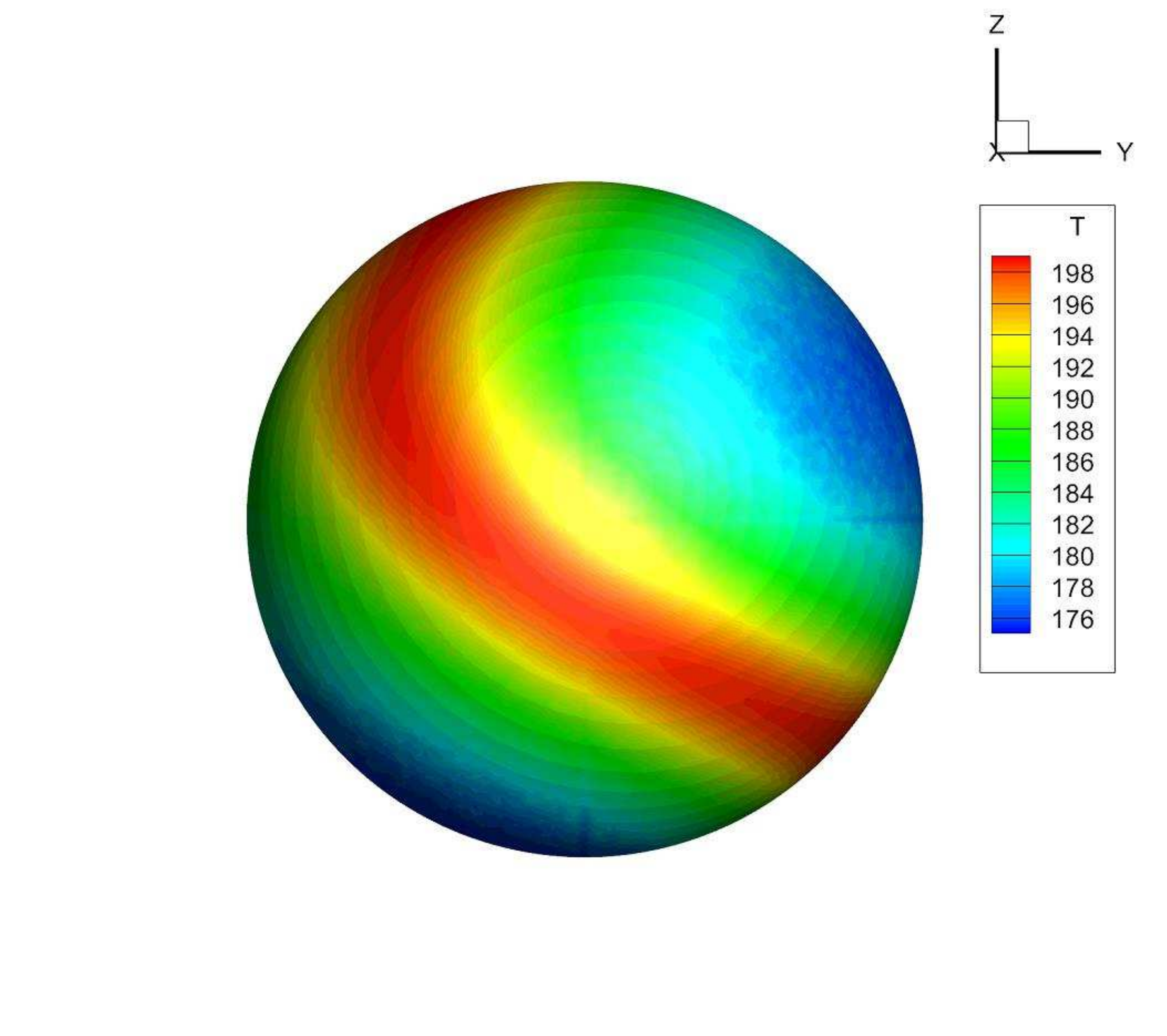}
	\caption{  Temperature distribution (in units of $10^6K$) in the NS crust for quadrupolar and dipolar magnetic fields with polar inductions $B_{pq} = 5\cdot10^{12}\,$G and $B_{pd} = 10^{13}$G correspondingly ($\beta = 0.5$). Magnetic axes are rotated from each other to the angle $\Theta_b  = \pi/4$, hereinafter a dipole axis is rotated on plots. The core temperature is $T_{core} = 2\cdot10^8K$. Left picture - cross-section in Z-Y plane, right one - NS crust surface.}
	\label{fig:Tcrust}
\end{figure}

In Fig.~\ref{fig:Tcrust} the temperature distribution of the NS crust is shown for magnetic dipole and quadrupole, rotated on an angle $\Theta_b  = \pi/4$ from each other, and the quadrupolar strength is a half from the dipolar one. The crustal temperature distribution is inverted in comparison to the surface one, i.e. the crust temperature is smaller in regions, where magnetic field is radial, and larger in the regions, where the field is almost tangential. The reason for this is the following. The heat flux is suppressed most crucially in the envelope, where the parameter $\omega\tau \gg 1$. The suppressed heat flux from the envelope in the NS regions with the tangential field (equatorial regions) causes the decrease of the temperature gradient in the crust. Thus, a variation of the crust temperature on the magnetic poles, where the field lines are radial, is higher, than on the equator (effect of heat blanketing envelopes). The temperature difference in the crust is about 10\% of its value.

\begin{figure}[!htp]
	\centering
	\includegraphics[width=7.5cm,height=7.0cm]{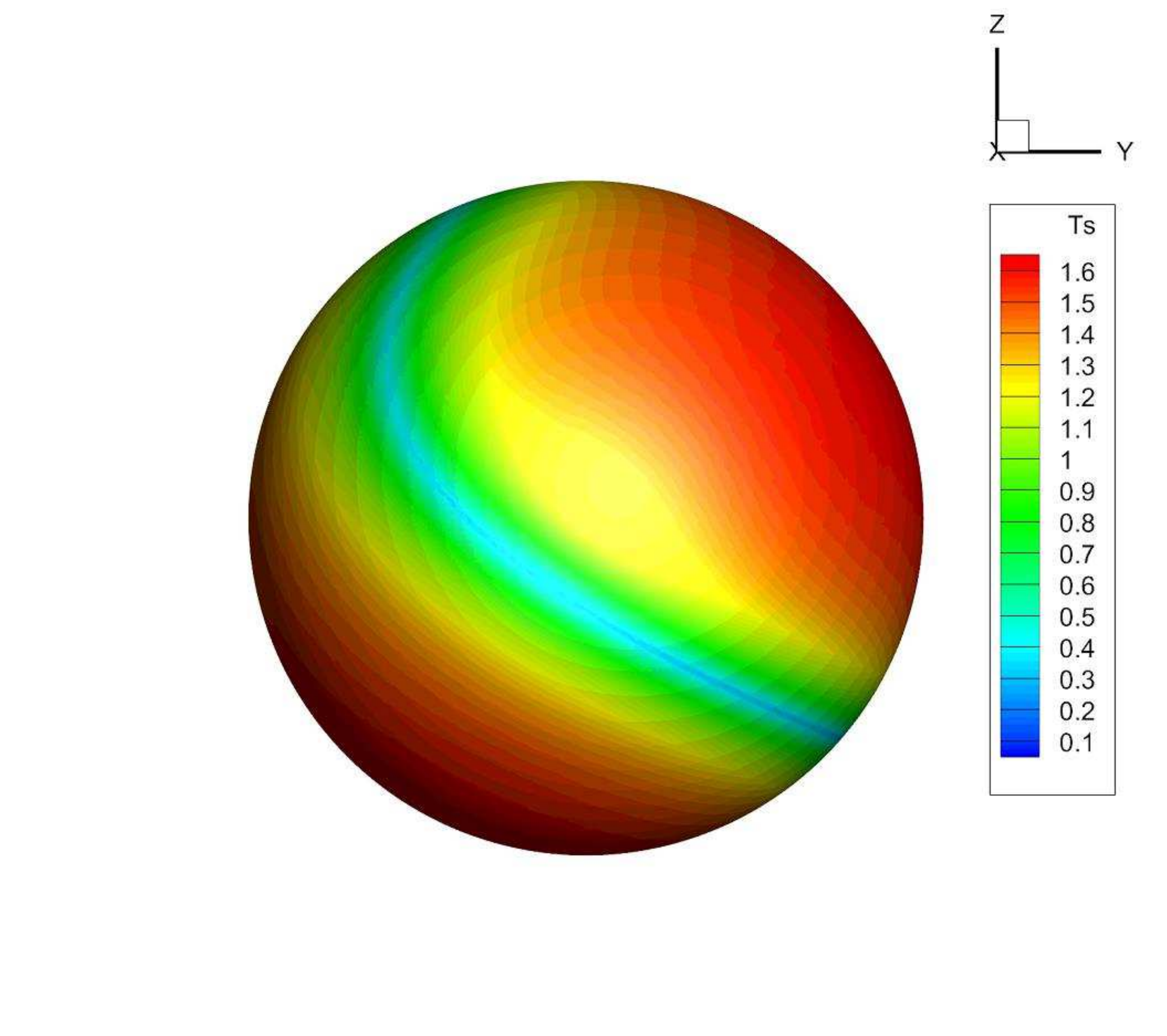}
	\includegraphics[width=7.5cm,height=7.0cm]{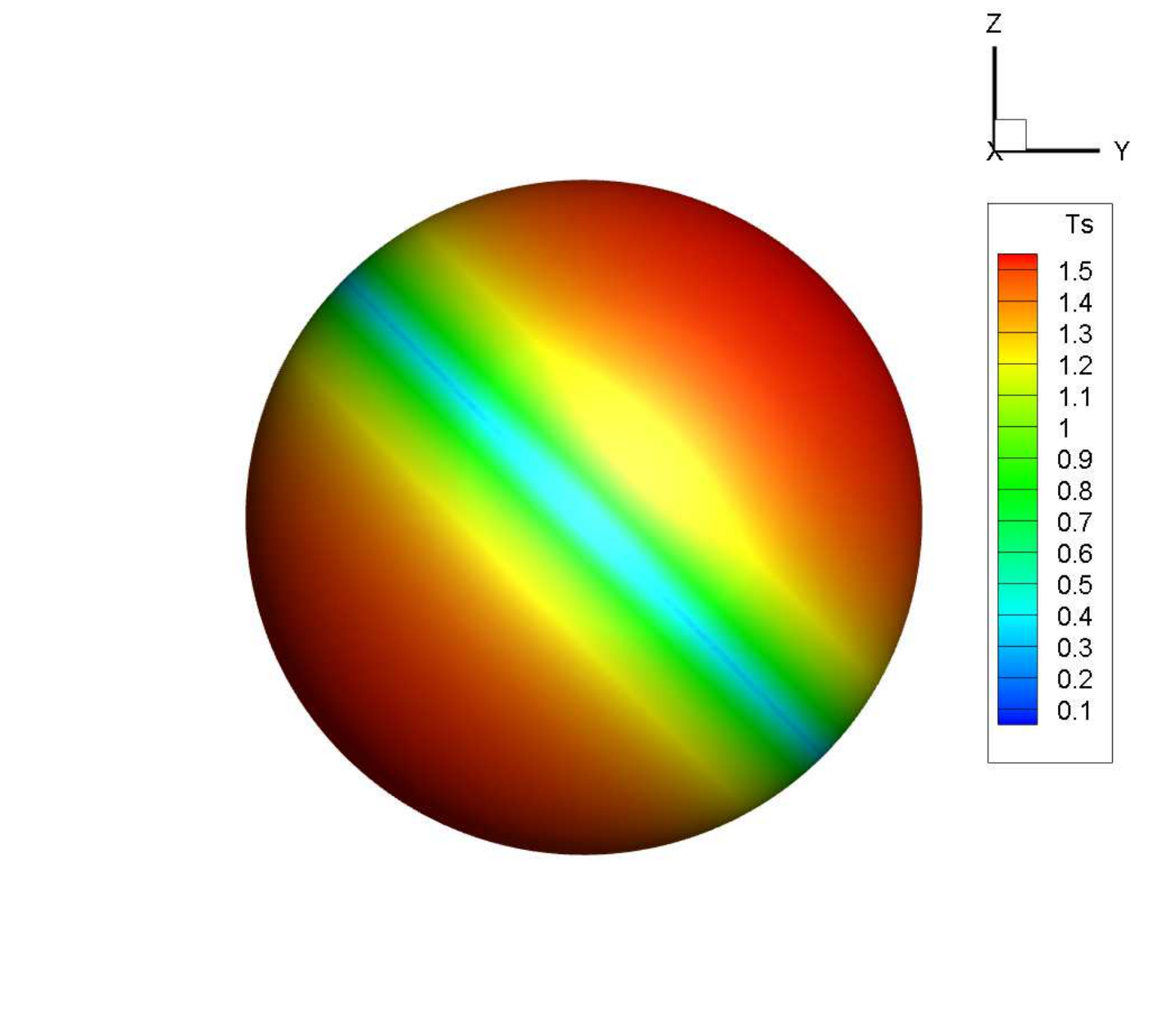}
	\caption{Surface temperature distribution (in units of $10^6K$) for the same parameters as on Fig.\ref{fig:Tcrust} (left panel) and for the NS without the quadrupolar field (right panel).}
	\label{fig:Tsurf1}
\end{figure}
The surface temperature distribution, that corresponds to the crustal temperature from Fig.~\ref{fig:Tcrust} is shown in Fig.~\ref{fig:Tsurf1} (left) together with the surface temperature of a NS with only a pure-dipolar magnetic field \cite{kondmoisBKGl} (right). The minimum temperature is approximately $3\cdot10^5K$, and the maximum one is near $1.6\cdot10^6K$. In a pure dipole case the surface temperature distribution is represented by two hot polar caps and a cold ring-shaped "belt".

 A "switching on" of the quadrupolar field effects on the heat transfer as follows. If the parameters $\beta<\sim1$ and $\Theta_b \neq 0$, the belt shape becomes irregular, and also the belt broadens from one side in comparison to the pure-dipolar configuration. Existence of the quadrupolar field causes the slight decrease of an effective temperature of a NS, so that the cold region is larger, if the quadrupolar component in the magnetic field is not negligibly small.

\subsection{Synthetic light curves}
Modelling of thermal light curves from the XDINSs is a well studied topic. A thermal emission from compact object was considered by a long list of authors (e.g. \cite{p1995, ps1996, zaneturolla,turolla13} with taking into account the effects of general relativity). For compact objects general relativistic effects may be significant. A rigorous relativistic theory of a light propagation near the compact object was developed in \cite{pfc1983}. In real conditions of the NS the effects of the general relativity are pronounced mostly by a redshift of the photon energy and a deviation of the photon's trajectory from the straight line, so that more than a hemisphere is observable, and an effective visible NS radius is more, than the exact one. We calculated a series of light curves, adopting a simple composite blackbody model, as it is described in \cite{p1995}. Bending angles are calculated adopting a simple analytical fit, proposed by Beloborodov \cite{bel2002}.

\begin{figure}[!htp]
	\centering
	\includegraphics[width=15cm,height=9.5cm]{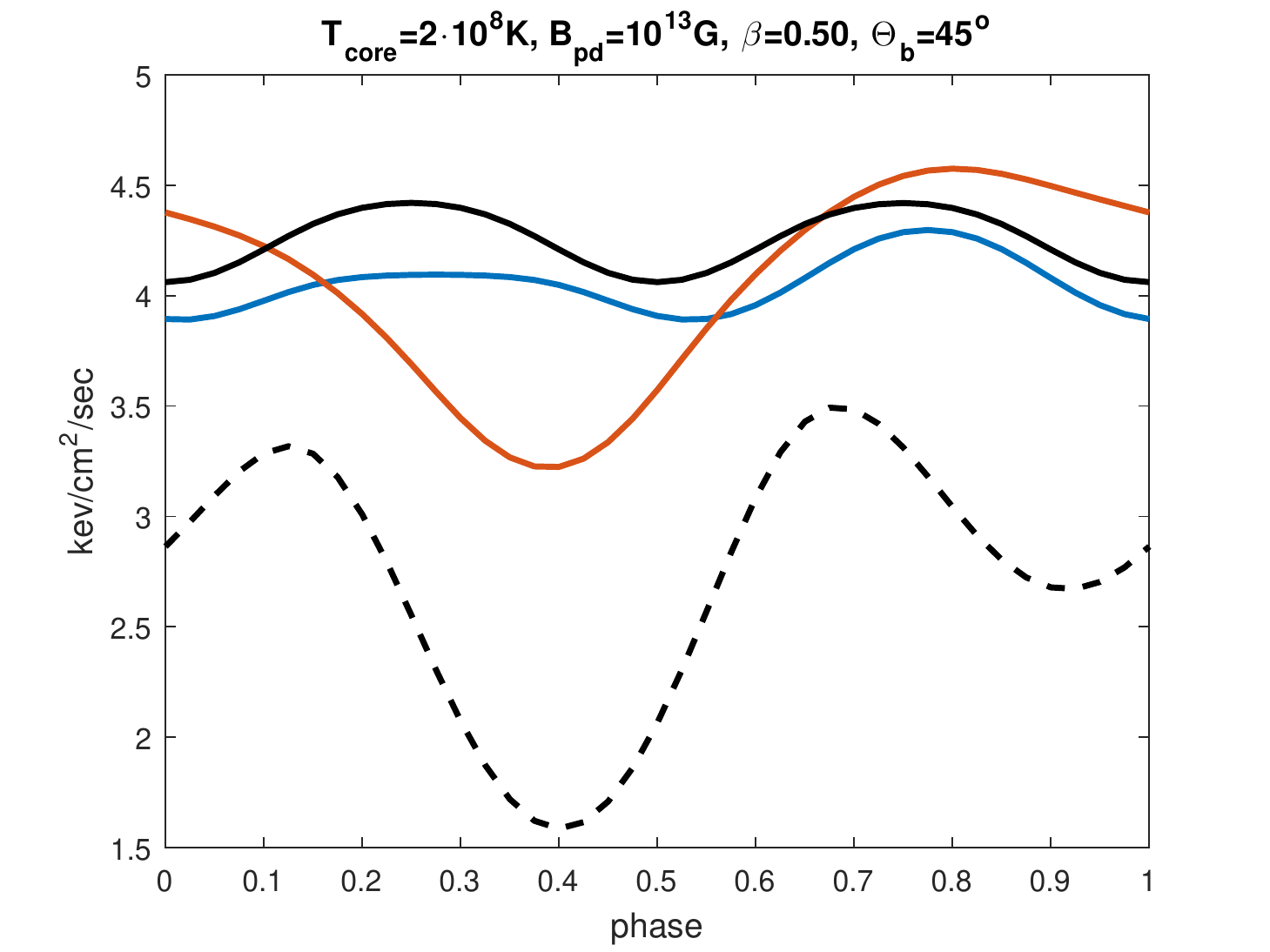}
	\caption{Light curves (for the energy flux) for the temperature distribution from Fig.\ref{fig:Tsurf1} for an orthogonal rotator ($\alpha_d = \zeta = \pi/2$). Blue lines correspond to the case, when rotational, dipolar and quadrupolar axes are in the same plane, red ones correspond to the case, when $\alpha_q = \alpha_d$, so both magnetic axes are visible for the observer. Black lines correspond to the light curves from the NS with a purely dipolar field. A dashed black line is the light curve from the Case $\alpha_q = \alpha_d$, but without taking into account light bending effects.}
	\label{fig:lc2p}
\end{figure}

In the absence of the quadrupolar component, the pulse profile is symmetric and sinusoidal, and light curve can be either two-peaked (both magnetic poles are visible) or one-peaked (one precessing pole is visible), and it is characterized by two angles: an angle between the rotational axis and the dipolar one $\alpha_d$, and the angle between the rotational axis and the line of sight to the observer $\zeta$. Quadrupolar component adds one more degree of freedom in a space of positions of the axes, which characterize the light curve, so it makes the analysis more complicated. We consider only two limits: when all three axes (rotational, dipolar and quadrupolar ones) are in the same plane, and when both dipolar and quadrupolar axes aligned with the line of sight to the observer in some moments of time during rotation, so that $\alpha_d = \alpha_q$, where $\alpha_q$ is an angle between the rotational and quadrupolar axes (e.g. when magnetic axes are in the equatorial plane with respect to the rotational one). We restrict ourselves with a constraint $\alpha_d = \zeta$.

Fig.\ref{fig:lc2p} shows the light curves for an orthogonal rotator ($\alpha_d = \zeta = \pi/2$), calculated for the temperature distribution from the previous subsection for both limits of the positions of the axes and for the pure-dipolar magnetic field configuration (black line). When all three axes are in the same plane (blue line), the light curve changes slightly from the dipolar one. One peak becomes narrower, and the second one is broadened in comparison to the pure-dipolar light curve, and the pulsations decrease slightly. An interesting situation occurs for the case, when $\alpha_d = \alpha_q$ (red line): the symmetry of the pulse profiles is broken, and light curves can have various non-sinusoidal shapes. Moreover, the pulsations are amplified sufficiently, from 4\% up to 14\% in comparison to the pure-dipole. Also it should be mentioned here, that the general relativistic effects can also change the pulse profile. The difference is, that in the absence of the light bending the light curve is two-peaked, while the general relativity effects make the pulse profile be one-peaked.

\section*{Acknowledgements}
This work was partially supported by RFBR grants 18-02-00619, 18-29-21021 and by the Program of the Russian Academy of Sciences P-12.

\end{document}